\documentclass[twocolumn,showpacs,preprintnumbers,amsmath,amssymb]{revtex4}

\usepackage{graphicx}
\usepackage{dcolumn}
\usepackage{bm}

\begin{document}

\preprint{APS/123-QED}

\title{The forget-remember mechanism for $2$-state spreading}

\author{J. Gu}
 \altaffiliation{}\email{guj@iopp.ccnu.edu.cn}
\author{X. Cai}%
 \email{xcai@mail.ccnu.edu.cn}
\affiliation{%
Complexity Science Center and Institute of Particle Physics,
Central China Normal University, Wuhan 430079, China
}%

\date{\today}

\begin{abstract}
A new mechanism, the forget-remember mechanism, is proposed for
studying the spreading process in 2-state model. Such mechanism
exhibits behaviors of message spreading influenced by some kinds
of functions about time and history caring about the individuals
of the spreading system, holding message or being out of message.
To demonstrate the mechanism, both linear and exponential forms
for forget-function and remember-function are simulated and show
that a great impact on the saturation of message-spreading and the
relative phase transformation.
\end{abstract}

\pacs{87.23.Ge,89.70.+c,87.23.Kg}
\maketitle

In this Letter we suggest a new mechanism, the mechanism with
forget and remember functions, to study the message spreading in
2-state model and show their behaviors impressed on spreading
process. Recent researches has widely addressed in the epidemic
modelling\cite{s1,s2,s3,s4,s5,s6,s7} and the biological
interaction, which has been shown valuable insights on
spreading\cite{s8}. Various epidemiological models has been
attracted attention of epidemiologists\cite{s9} and it was pointed
out that epidemiological processes can be regarded as
percolation\cite{s5,s10,s11,s12}.

As known, most of spreading systems should be considered to have
the following stronger resemblances: (i) In any of the spreading
systems, there are certainly some special objects to be
transmitted, which could be classified into two categories, things
and messages. Things are not like messages. Things in a spreading
system should be conservative, e.g. passengers on a flight were
transported on airport network from City-A to arrive City-B, they
are no longer in City A\cite{s13,s14}. But messages can be copied
with some freedom and without conservation of the total number,
e.g. a news you delivered to your friend, you still hold the news.
(ii) Messages not only might be spread or lost, but also might be
forgotten(not completely lost) or remember(after forgotten). (iii)
Messages could be multifarious, such as virus on internet, e-mail,
rumors in public, epidemic forest fire and so on\cite{s9,s15}.
(iv) Each individual inside a message-spreading system could be at
one of two states(holding message or being out of message) in
2-state models. And there still exist two cases, in the first
case, an individual could transmit message to others or not, and
in the second case, an individual could get message infected from
the others or not. Altogether four states. No more other states
for message-spreading systems. For example, the SI model assumes
that each individual must be in one of two possible states, either
susceptible(without message and can be transmitted in) or
infective(with message and can be transmitted out). The SIR model
adds the third state\cite{s16,s17}, removed state(lost message but
can not be accepted message again). (v) Individual copies message
into or from its neighbors, topologically only, not regarding the
spacial distance and the limited transmitting speed between
individuals usually. (vi) Time of evolution of the spreading
system is sometimes considered discrete replacing continuous.

Based on the above-mentioned understanding which we prefer to call
it the message theory to information theory, we will focus our
attention on the influence of a mechanism, the forget-remember
mechanism(FRM), in our investigation on message spreading. As we
know, in the most of ready-made spreading studies, the ordinary
means assumed a constant to measure the probability of spreading,
and the most important parameter expressing the effective spreading
rate, which not only to decide the percent of individuals with
message, but also to determine whether the prevalence can break
out\cite{s16,s17}. It is not a proper supposition here. Message in
spreading maybe resemble knowledge in learning. In nature, the
learning curve discovered by German Experimental Psychologist
Hermann Ebbinghaus is famous\cite{s18,s19}, which could be
corresponding in spreading dynamics. Our so-called FRM is described
now as following. Firstly, when an individual holds message, it may
lose it with probability $P_{-}(t)$, a function on time $t$. We
ponder a circumstance that the longer the individual holding
message, the easier it forget message. Secondly, when an individual
had seized message but lose it later, it can recall that message
easier after shorter time with probability $P_{+}(t)$, also a
function on time $t$. Taking notice, here the time in $P_{\mp}(t)$
in the FRM is not same to the system-time, but only cares about the
history of each individual. The time in forget mechanism is
zero-counting at the moment when the individual got message, while
in remember mechanism, zero-counting at the moment when it lost
message. We will show the influence of the FRM on time-interval and
how it to impact on spreading curve.
\begin{figure*}
\includegraphics{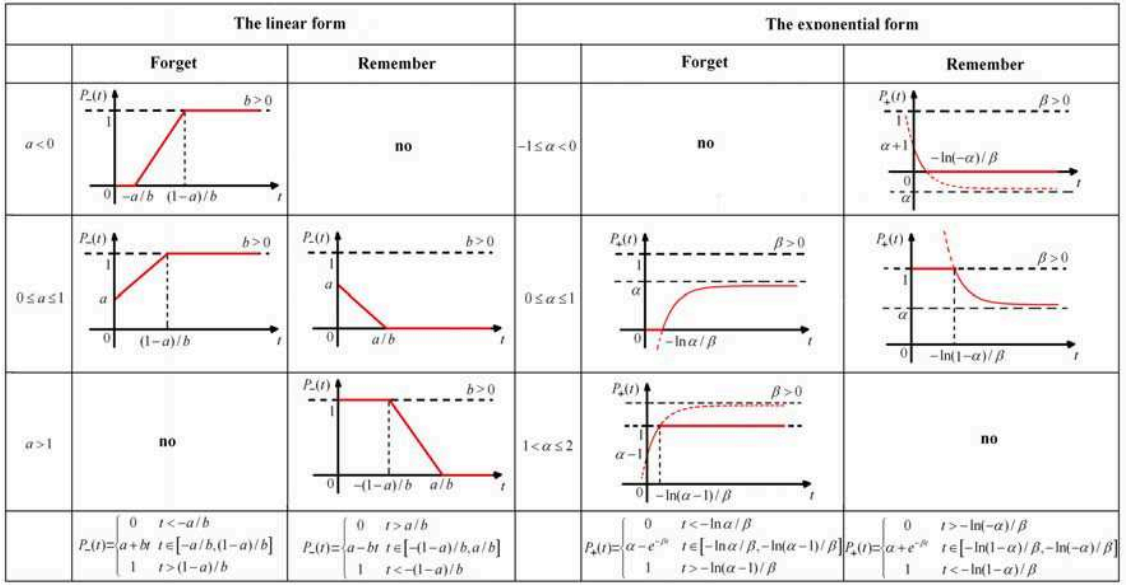}
\caption{\label{fig:wide}Functions of the FRM: the linear form and
the exponential form. $P_{-}(t)$ is probability for the forget
mechanism while $P_{+}(t)$ for the remember mechanism, both are
ranged from zero to one.}
\end{figure*}

\begin{figure}
\includegraphics[width=9.1cm]{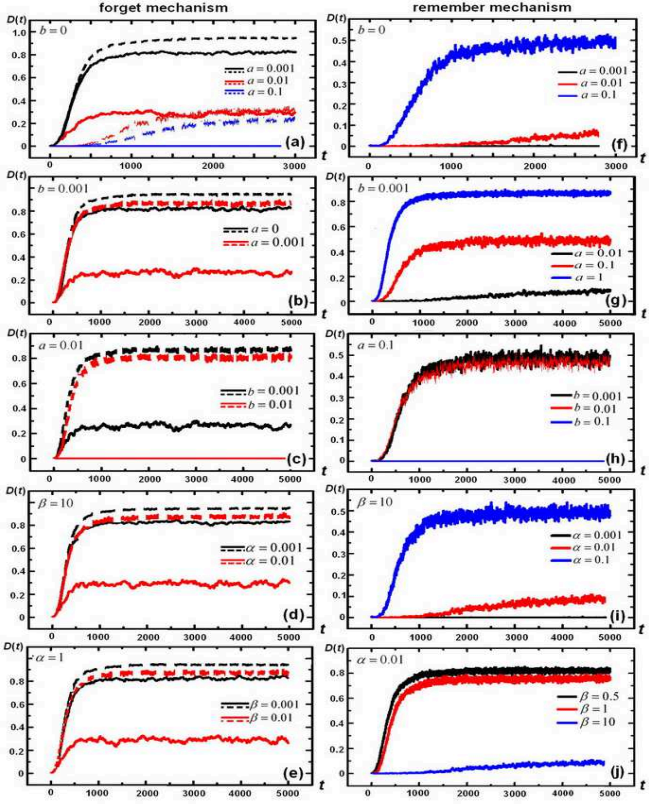}
\caption{\label{fig:epsart} Plot of the density $D(t)$ versus the
system-time step $t$ with the spreading probability $\nu=0.002$.
Figures on the left-row for forget mechanism and on the right-row
for remember mechanism. The dashed curves are simulated with fixed
remember probability $P_{+}=0.1$ while the solid curves without
remember mechanism in left figures. Figures on the upper three
lines by using linear form with values of parameters $a$ and $b$,
and those on the lower two lines by using exponential form with
values of parameter $\alpha$ and $\beta$ as shown.}
\end{figure}

\begin{figure}
\includegraphics[width=9.1cm]{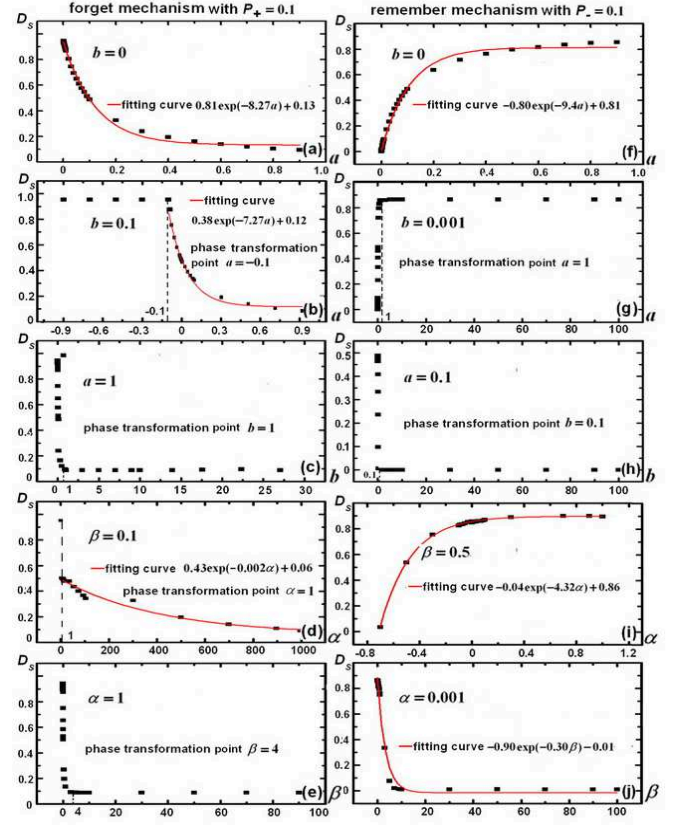}
\caption{\label{fig:epsart} Saturation density $D_{s}$ versus
parameters in different forms. Figures on the left-row for forget
mechanism and on the right-row for remember mechanism. Figures on
the upper three lines by using the linear form with $a$ and $b$,
and those on the lower two lines by using the exponential form
with $\alpha$ and $\beta$ as shown.}
\end{figure}

We attempt to consider our FRM  by using a developed model, e.g.
the Barab$\acute{{\rm a}}$si-Albert(BA) algorithm as a frame for
message-spreading, because it can reflect the social network
strikingly and its degree distribution exhibits a power-law
behavior of with form $P(k)\thicksim k^{-\gamma}$ where exponent
$\gamma$ ranged between 2 and 3\cite{s15,s20}. In
Ref.\cite{s9,s20,s21} the epidemic processes in an uncorrelated
network do not possess an epidemic threshold, and below which the
diseases cannot produce a macroscopic epidemic outbreak. In a
spread process, if an infected individual transmits message with
probability $\delta$, and a susceptible individual is infected by
its infected neighbor with probability $\nu$, the proportion of
$\nu$ and $\delta$ is defined as an effective spreading rate
$\lambda$. Taking the BA networks, and let $\rho_{k}(t)$ the
probability of a message-holding individual with degree $k$, we
get an equation for the mean field function
\begin{equation}
{\frac{\partial\rho_{k}(t)}{\partial t}=-\rho_{k}(t)+\lambda
k[1-\rho_{k}(t)]\Theta(\rho_{k}(t))}
\end{equation}
where $\Theta(\rho_{k}(t))$ is the probability of any given link
to a message-holding individual\cite{s9}. Then we can reach a
spread threshold $\lambda_{c}={\langle k\rangle}/{\langle
k^{2}\rangle}$, so that $\lambda_{c}\rightarrow 0$ if amount of
individuals in a network $N\rightarrow\infty$\cite{s22,s23}.

Choosing the BA networks built with the preferential attachment
character\cite{s15} with amount $N=1000$. To beginning with an
initial network with $m_{0}=5$ individuals, a new individual $i$
is added to the network linked to an existing individual $j$ with
probability: $\Pi_{i}={k_{i}}/{\sum_{j}k_{j}}$, where $k_{i}$ and
$k_{j}$ are degrees of individuals $i$ and $j$. The average degree
$\langle k\rangle=3.8$ and the exponent $\gamma=2.2$. Now let us
address the forget mechanism. We assume that an individual without
message on the network can get message from its neighbors which
are holding message with probability $\nu$. An individual which is
in a message-holding state may become into a message-losing state
with the probability function $P_{-}(t)$, where concerning the
time $t$, i.e. how long this individual holding message.
Certainly, $t>0$ and the values of $P_{-}(t)$ range from zero to
one. Meanwhile the remember mechanism is built by considering an
individual in a message-losing state, but it can recall message
only if it had possessed of this message before. This case is
common in natural, especially it takes place on those processes,
e.g. language-learning, disease-catching or rumors-circulating
repeatedly and so on. An individual at a state without message,
but which must formerly have gotten message by copying or infected
from its neighbors, can remember message controlled by the
probability function $P_{+}(t)$. Usually the longer the individual
was out of the message, the harder it remember the message. The
ability to recall the past should be concerning the history, that
is to say, if an individual had never owned message and then had
never undergone on a forget-process, it can not remember any
message. Time $t$ in $P_{+}(t)$ is zero-counting when it lost
message for each individual respectively.

In order to gain further insight into the FRM, we consider two
forms for both $P_{-}(t)$ and $P_{+}(t)$: the linear form and the
exponential form, see Eq.(2) with parameters $a$ and $b$ and
Eq.(3) with $\alpha$ and $\beta$. To make our mechanism clear,
parameter $b$ or $\beta$ had better take more than zero to decide
the shapes of probability functions for forget and remember
mechanisms with additional constant terms $a$ or $\alpha$. By the
way, when $b$ or $\beta$ is equal to zero, i.e. the probability
functions should be simply constant, we shall discuss the first
form only.

The linear function is defined as
\begin{equation}
 P_{\mp}(t)=\begin{cases}
0 & \mbox{\ \rm for\ \ } \pm t<-a/b, \cr a\pm bt &\mbox{\ \rm for\ \
} \pm t\in[-a/b,(1-a)/b], \cr 1 &\mbox{\ \rm for\ \ } \pm t
>(1-a)/b.
\end{cases}
\end{equation}
and the exponential function as
\begin{equation}
 P_{\mp}(t)=\begin{cases}
0 & \mbox{\ \rm for\ \ } \mp e^{-\beta t}>-\alpha, \cr \alpha\mp
e^{-\beta t} &\mbox{\ \rm for\ \ } \mp e^{-\beta
t}\in[1-\alpha,-\alpha], \cr 1 &\mbox{\ \rm for\ \ } \mp e^{-\beta
t}<1-\alpha.
\end{cases}
\end{equation}

Since $P_{\mp}(t)$ are probabilities for forget and remember
functions, they are ranged between 0 and 1 as shown in Fig.1. The
behavior of the influence of the above different forms to the
spreading system can be computed by noticing that the
zero-counting of time $t$ at which each individual forgets or
remembers message separately. let $D(t)$ be the density of
individuals with message on the frame in the BA model. As showed
most obviously in Fig.2, the influences make a faster progress to
the spreading process in the early time-steps of the evolution,
and always reach saturation, i.e. not converting whole individuals
of the system after long time spreading, both in linear and
exponential forms.

Not matter how many the parameters were taken in both forms with
the forget mechanism, the value of $D(t)$ in the saturation is
descending with parameters increasing. And with the remember
mechanism, since the value of function is ascending with
parameters $a$ and $\alpha$ increasing, the $D(t)$ at the
saturated moment is also rising with them, but is opposite with
the parameters $b$ and $\beta$. As shown, in the left five figures
of Fig.2, the dashed curves are considered the forget mechanism
with fixed remember probability while the solid curves without
remember mechanism. The results show that the density $D(t)$ in
the case of the left figures might be easy saturated. The reason
is that the remember mechanism inspirits the prevalence of
message, while the forget mechanism restrains the spreading.

Let $D_{s}$ be the saturated value of $D(t)$ in different cases.
Fig.3 exhibits the relation between $D_{s}$ and parameters $a$,
$b$ and $\alpha$, $\beta$. Noteworthily, as shown in Fig.3 the
saturation density $D_{s}$ can vary in a very large range (from 0
to 1) controlled by parameter taking. That is to say, the FRM is
not a slimly adjustment on the spreading process, which is to the
contrary to the conclusion that only the effective spreading rate
is important to the spreading process in some fashionable models.
In our figures, part of the relations between the saturation
density $D_{s}$ and parameters could be fitted by the exponential
function. And as shown, there exists the thrill of a new sight
which is no other than the phase transformation points about the
saturation density $D_{s}$. At the phase transformation point, the
data are divided into two sides, on the one side, $D_{s}$ is
ascending or descending quickly, and on the other side, $D_{s}$
changes very slowly. This observation based on the FRM is
different from the present study of the spreading process on the
BA network\cite{s20,s21}.

In summary, the FRM we have presented for spreading process with
2-state systems is possibly the simplest one in a kind of selected
practical networks(the BA model) currently in effect. A novel
feature in the model with our forget-remember mechanism is the
saturation of the spreading density with phase transition
controlled by the parameters, i.e. the forget coefficients and the
remember coefficients. The results can be concluded as follow: (i)
The probability in spreading process should not be always
constant, maybe  depends on time $t$, especially cares about the
history of each individual holding the message. (ii) Evidently the
forget mechanism restrains the spreading but the remember
mechanism inspirits it. Truth should be simple and obvious. (iii)
The effective spreading rate is not the only possible important
factor to decide the density of the message-holding individuals
and whether the prevalence can break out. The FRM would naturally
work on the spreading systems to let the spreading process reach
the saturation with varying value ranged widely from zero to one
at most. It serves as a great source of inspiration for the
spreading process control. (iv) There exists the phase
transformation on the saturation of spreading density. It needs to
investigate further.

The FRM on the spreading system can explain why different diseases
have different saturation in population, even though they have the
semblance effective spreading rate. Different attitudes derive
from different perspectives. It is sometimes better to emphasize
the attention to history of each individual inside the system.
Details is the key of characteristic of the whole.

This work is supported by the National Natural Science Foundation
of China (Grant Nos. 70571027, 10647125,10635020 and 70401020) and
the Ministry of Education of China(Grant No. 306022).

\end{document}